# Detection of cancer stages through fractal dimension analysis of tissue microarrays (TMA) via optical transmission microscopy

Liam Elkington, Prakash Adhikari, and Prabhakar Pradhan*

[1]Department of Physics and Astronomy, Mississippi State University, Mississippi State, MS, USA, 39762

*PPradhan: pp838@msstate.edu

**Abstract:** Tissues are fractal due to its self-similar structure, and the fractal dimension change with the abnormalities such as in disease like cancer. The optical imaging of thin slices of tissue using transmission microscopy can produce an intensity distribution pattern proportional to its refractive index which represents the mass density distribution pattern of the tissues. The fractal dimension of tissue is calculated by analyzing this mass density distribution and is known to increase with the progression of carcinogenesis. This paper explores the viability of using this quantitative approach of fractal dimension analysis to create a standardized accurate cancer diagnosis test and staging which reduces the issues plaguing current testing methods. A commonality for most of the deadliest cancers is their lethality due to the difficulty in properly diagnosing them in the early stages or accurete late staging. This difficulty can arise from the physical location of the organ causing them to be hard to access and the lack of noticeable symptoms until the late stages when there is almost no hope of treatment. To study some of these deadly cancers, commercially available paraffin embedded tissue microarray (TMA) samples containing multiple cores of different cases and stages of Pancreatic, Breast, Colon, and Prostate cancer are analyzed. The fractal dimension of different TMA samples is able to correctly differentiate between the different stages of each cancer, raising the possibility of a standardized system being created to increase diagnosis accuracy in the future.

**1. Introduction**

With millions of people dying from cancer each year according to the WHO, the accurate and early detection of cancer is necessary for preventing this widescale mortality (1). Cancer arises in people due to changes in the DNA replication process that occur during cell reproduction (2). Cancerous cells tend to reproduce themselves more rapidly than normal cells, causing a massive buildup of abnormal cells known as a tumor. These tumors inhibit the normal bodily functions of the organs they reside in, resulting in a potentially fatal outcome if left to grow unchecked. In particular, cancer cells metastasize to other vital organs and strat growing those secondary places. Currently, most cancer tests rely on a pathologist qualitatively studying a sample from a biopsy for minute structural differences or on chemical methods involving and time-consuming dyes and stains (3). Both of these methods are inefficient and potentially inaccurate due to the length of time needed to utilize the dyes and the reliance on the observational abilities of the pathologist. Due to high potential for human error in conventional techniques, a more quantitative, mathematical test is necessary.

Quantitative approaches in detecting structural change with the progression of carcinogenesis at the earliest possible point are in high demand. Tissues have spatial heterogeneity in their mass density distribution and a self-similar structure. This self-similar structure can be analyzed and expressed in terms of fractal dimension. The fractal dimension of an object is a number quantifying how similar the structure remains with changes in length scale. A tissue's fractal dimension will change throughout the course of cancer due to the increased production and rearrangement of intracellular structures such as DNA, RNA, lipids, heterochromatin and the extracellular matrix which causes an increase in the mass density of the tissue. Because cancer affects a tissue's fractal dimension, a quantitative diagnosis test can be developed based on these changes (4,5). To test the viability of this method of diagnosis, tissue microarray (TMA) samples were utilized. TMA samples are a method of commercially available clinical sample observation that is increasingly growing in popularity. They consist of a glass slide containing multiple cores of consistent paraffin embedded tissue samples from the same bodily region of several people with varying stages of cancer. The cores consist of a circular sample of tissue that is 1.5 mm in diameter and 5 μm in thickness embedded in paraffin wax. Each slide can contain up to several hundred cores, greatly increasing the speed and high throughput analysis at which tissue samples can be processed and analyzed (6,7). For the purpose of this experiment, four different cancer TMA slides containing only 24 cores each of different cases were used. The types of cancer studied were Pancreatic, Breast, Colon, and Prostrate. Pancreatic cancer was chosen because of its lethality, it is estimated that 45,750 people will die from this disease in the United States in 2019 (8). Its mortality is due in large part to the organ's location being difficult to reach and because it does not display any prominent chemical changes or symptoms until the hard to treat late stages. Another cancer that was studied was Breast cancer due to its prevalence. In 2019, it is estimated that over 268,000 women will be diagnosed with Breast Cancer which is characterized by a diverse range of potential causes including radiation exposure, consumption of alcohol, and age at which a first pregnancy occurs (8,9). Colon and Prostate cancer, both of which have been linked to high consumption of red meat, also exhibit high rates of mortality and diagnosis rates so they too were studied using the fractal dimension technique (8,10,11). It is estimated that in 2019 these four cancers will be responsible for the deaths of over 170,000 people within the United States alone (8). Because of either their prevalence, lethality, or combination of the two Pancreatic, Breast, Colon, and Prostate cancer were analyzed using this quantitative method which has the potential to become a new diagnosis method.

## 2. Methods

### 2.1. Mathematical Methods

#### 2.1.1. Fractals and Fractal Dimensions

Fractals are structures that exhibit self-similarity at different length scales and can be divided into two categories: random and deterministic. Deterministic fractals are generated in a purposeful way while random fractals are created through a stochastic process (12). Random fractals occur quite often in nature and can be seen in the structures of trees, coastlines, and most importantly for this experiment in tissue structure. These random fractals can have varying degrees of self-similarity which leads to the need for a quantifiable fractal dimension. A structure's fractal dimension is essentially a measure of how self-similar the structure is, the higher the dimension, the more self-similar it is. The fractal dimension of a random fractal is determined using a box-counting method. This method works by first placing the fractal structure on an evenly spaced grid and counting the number of boxes needed to cover the structure (13,14). The fractal dimension is then calculated by refining the grid through a box-counting algorithm. The equation used in this algorithm is as follos:

$$D_\text{f} = \frac{\ln(N(r))}{\ln\frac{1}{r}}$$

where N($r$) is the number of boxes of side length $r$ needed to cover the fractal structure. $D_f$ is the average fractal dimension of the structure since the box counting method accounts for several different realizations. $D_f$ is calculated from a ln(N($r$)) vs lin(1/$r$) slope by varing length scale $r$.

*2.1.2 Calculating Fractal Dimension using Microscopic images.*

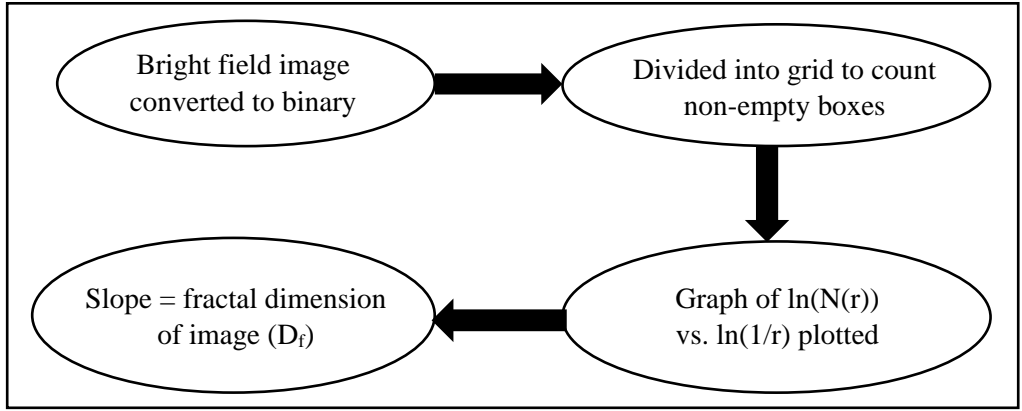

Fig. 1: The step by step process for calculating the fractal dimension of a microscopic image is outlined.

This method has already been utilized in studies on several different types of cancer (4,15). It works by taking the binary version of an image and applying the popular box-counting method to divide the image into a grid of evenly sized boxes and creating a graph of ln(N(r)) vs. ln(1/r) whose slope is equal to the fractal dimension. While most of these experiments focused on a single type of cancer, the experiment in this paper deals with multiple types of cancer to explore the potential of creating a standardized numerical index that applies to multiple types of cancer.

**2.2. Physical Set Up**

*2.2.1 TMA Samples*

Four different TMA samples from US Biomax were studied in this experiment: Pancreatic (T142b), Breast (BR248a), Colon (T054c), and Prostate (T191a). Each slide consists of 24 cores of 1.5 mm in diameter and 5 μm thick tissue that were embedded in paraffin wax. For each sample, at least two cores of each stage of cancer are analyzed, at the several parts of the core. Each core on the Pancreatic and Colon TMA slides coomes from individual patients of varying age and sex. The Prostate TMA slide contains 12 pairs of identical cores where each pair was obtained from a different patient. The Breast TMA slide contains individual cores from female patients of varying ages. These cores were either normal tissue samples or tissue samples in stage I, stage II, or stage III of cancer.

*2.2.2. Optical Microscope Setup and Imaging*

The images are taken using an Olympus BX61 Upright Bright field Microscope (Olympus, USA) coupled with an Amscope model MU1003 camera using Amscope software. Each TMA slide is imaged on the microscope stage which is being operated in transmission mode. At least 10 different cores are imaged on each slide depending on the number of cores and stages of cancer present. Each core is imaged with 50x magnification objectives without immersion oil in at least 10 different spots. Every spot has 5-10 images taken at slightly different focusing heights to ensure at least one focused image is captured within the working distance of the objective.

For every biological samples, certain refractive index properties can be attributed. These properties show a distinct change when cancer present in the sample causes a small linear increase in the sample's mass density (16). For this study, the contrast of the grayscale images taken with the microscope is assumed to be caused by the spatial mass density variation within the sample. This variation provides the refractive index variation which in turn provides the correlation between the mass density of the sample and the pixel intensity/transmission intensity of the image.

### 2.3. Analysis of Images

Each captured optical image is analyzed using *Image J* (NIH, USA) software. First the bright field images are converted into 8-bit binary images with *image J*. Then a smaller area that looks similar is selected for each realization, and its surroundings cropped to remove any dead space or deformities in the tissue structure of a TMA sample. Finally, the fractal dimension is calculated using the box-counting method which is performed by the fractal analysis through the method mentioned above in section 2.1.2.

## 3. Results

After collecting optical microscopic images of each sample, they are converted into binary images in *ImageJ*. Following that the fractal dimension analysis tool in *ImageJ* is used to calculate the fractal dimension of each cancer stage of TMA samples which are then ensemble averaged and bargraphed for comparisons. As seen in the graphs, the fractal dimension for the Pancreatic, Breast, Colon, and Prostate cancer samples increases with the progression of the cancer through the different stages. This increase in fractal dimension is due to the fact that the presence of cancer results in a higher cell replication rate (17). The rate of cancerous cell replication is greater than that of normal cells because of genetic mutations. These genetic mutations also cause the cancerous cells to spread to other regions of the body at a higher rate than the normal cells (18). As the cells reproduce at a faster rate, the mass density of the affected tissue area increases. Since the fractal dimension of a tissue sample is dependent on this mass density, it will increase as the mass density increases due to the progression of cancer.

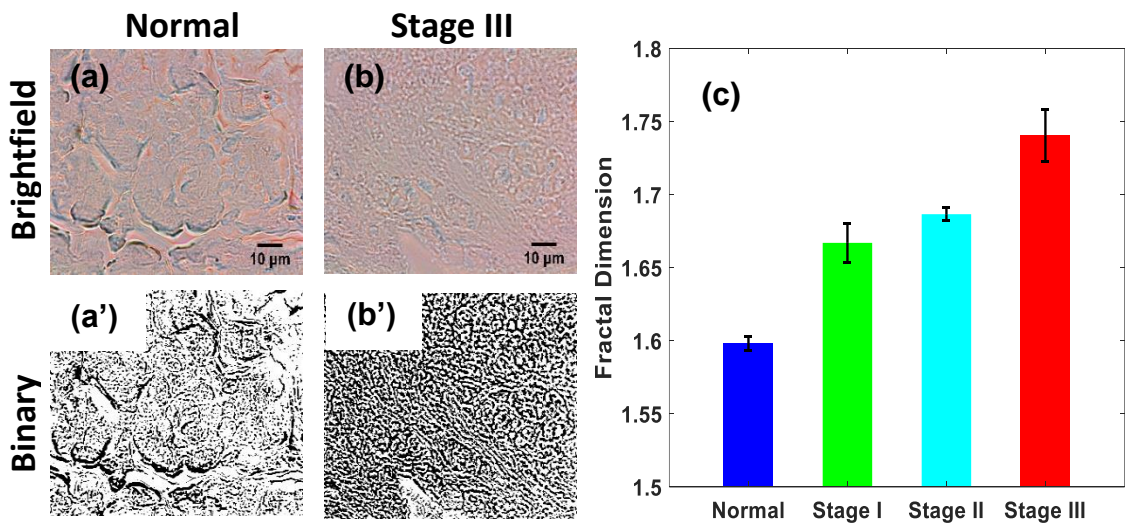

Fig. 2: (a) and (b) are the brightfield images of the normal and stage III Pancreatic TMA. (a') and (b') are the corresponding binary images. (c): The bar graph of the average fractal dimension of the pancreatic tissue samples (n=10). The results show fractal dimension of cancer stage I increases by 4%, stage II by 6%, and stage III by 9% with respect to the normal. (P<0.05)

Figs. 2 (a) and (b) show the brightfield images of normal and stage III Pancreatic tissue from a TMA sample respectively. Figs. 2 (a') and (b') show the binary versions of each brightfield images. These binary images clearly display the increased mass density distribution which results from the progression of cancer. As seen in Fig. 2 (c), the normal Pancreatic tissue sample had the lowest fractal dimension while the different stages of cancer possess increasing fractal dimensions. The actual fractal dimensions for each sample are calculated to be 1.5984 for the normal, 1.6673 for stage I, 1.6866 for stage II, and 1.7407 for stage III. The fractal dimension percent difference between the normal and stage I sample is 4%, between normal and stage II is 6%, and between normal and stage III is 9%. This percentage increment in the logarithm scale of the fractal dimension with the increase in the stage of cancer is highly significant. These results suggest that the fractal dimension of a Pancreatic tissue TMA sample increases with the progression of cancer. This is a logical outcome as cancer is known to cause cell replication to increase which then causes the higher density, and therefore the self-similarity, of cells within the tissue to increase as mentioned earlier. This results in variation to the transmission intensities from the sample which increases the fractal dimension. Since fractal dimension is a measure of the self-similarity of a sample, the fractal dimension of Pancreatic tissue will increase with the progress of cancer. Similar results have also been observed in several other studies (19).

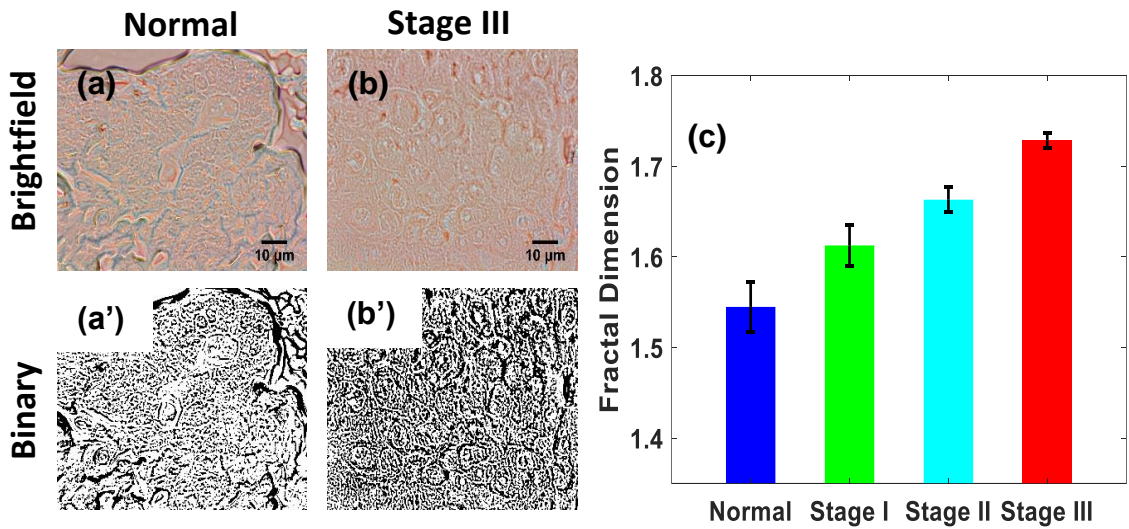

Fig. 3: (a) and (b) are the brightfield images of the normal and stage III Breast TMA. (a') and (b') are the corresponding binary images. (c): The bar graph of the average fractal dimension of the Breast cancer tissue microarrays (TMA) samples (n=10). The results show fractal dimension of cancer stage I increases by 4%, stage II by 7%, and stage III by 12% with respect to the normal. ($P<0.05$)

The brightfield images from a normal and stage III Breast cancer TMA tissue core are shown in Figs. 3 (a) and (b) respectively. Their binary analogs are shown in Figs. 3 (a') and (b') with (b') showing an increased density. The greater density demonstrates how the progression of cancer results in an increased mass density distribution. Shown in Fig. 3 (c), the normal Breast tissue sample has the lowest fractal dimension while the different stages of cancer had the fractal dimensions increasing with the progression of cancer. The actual fractal dimensions of the Breast tissue samples are 1.5448 for the normal, 1.6126 for stage I, 1.6631 for stage II, and 1.7284 for stage III cancer. For the normal and stage I tissue samples, the percent difference between the two is 4% while the percent difference between the normal and stage II and the normal and stage III are 7% and 12% respectively. These results suggest that the fractal dimension of a Breast TMA sample increases with the progression of cancer. Cancer is known to cause an increase in the replication rate of cells which therefore causes an increase in the tissue's mass density distribution. These results make sense based on this fact because the fractal dimension of a tissue sample is a measure of its mass density distribution. Similar results to these have also been observed in other studies (20).

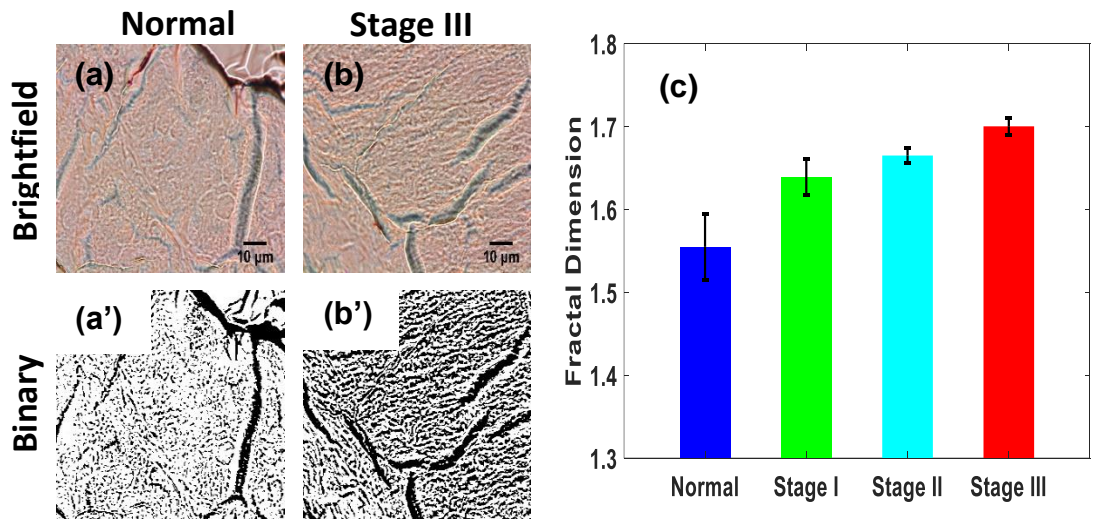

Fig. 4: (a) and (b) are the brightfield images of the normal and stage III Colon TMA. (a') and (b') are the corresponding binary images. (c): The bar graph of the average fractal dimension of the Colon cancer tissue microarrays (TMA) samples (n=10). The results show fractal dimension of cancer stage I increases by 5%, stage II by 7%, and stage III by 9% with respect to the normal. (P<0.05)

Normal and stage III cores of the Colon TMA slide's brightfield images are shown in Figs. 4 (a) and (b). Their corresponding binary images are pictured in Figs. 4 (a') and (b') and show how the mass density distribution of a tissue sample increases throughout the progression of cancer. As can be seen in Fig. 4 (c), the normal Colon tissue sample possesses the lowest fractal dimension while the fractal dimensions for each stage of cancer increases with its progression. The actual fractal dimension values for each sample are 1.5551 for the normal, 1.6393 for stage I, 1.6652 for stage II, and 1.7004 for stage III. When the normal tissue sample is compared with stage I, stage II, and stage III samples, the percent differences are found to be 5%, 7%, and 9% respectively. These results suggest that the fractal dimension of a Colon TMA sample increases with the progression of cancer. Like other tissue samples, the presence of Colon cancer causes the replication rate of the affected cells to increase resulting in an increased mass density distribution. The fractal dimension of the tissue sample will increase due to this, so the results gathered from this study make sense. Similar results have also been observed in other studies (8).

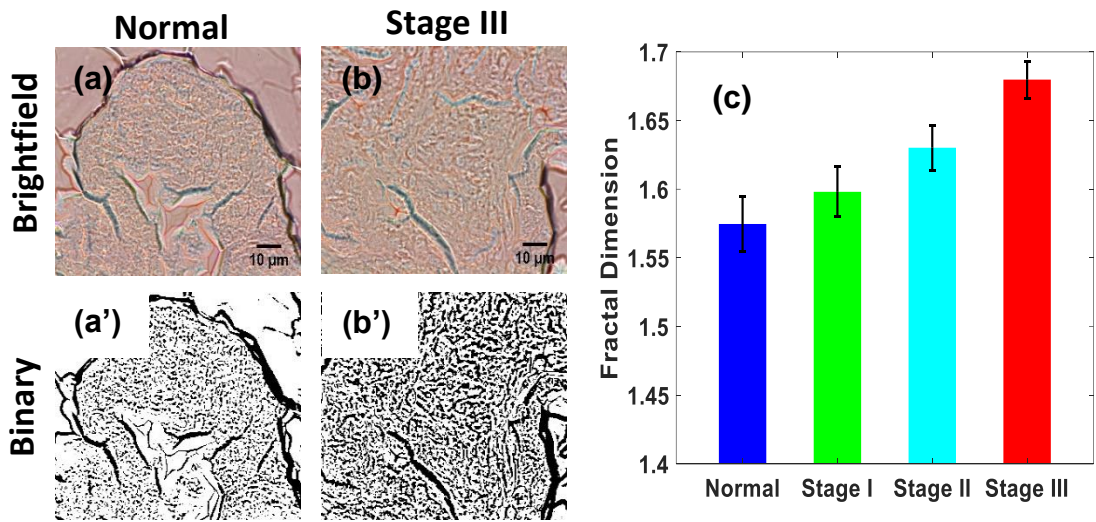

Fig. 5: (a) and (b) are the brightfield images of the normal and stage III Prostate TMA. (a') and (b') are the corresponding binary images. (c): The bar graph of the average fractal dimension of the Prostate cancer tissue microarrays (TMA) samples (n=10). The results show fractal dimension of cancer stage I increases by 2%, stage II increases by 4%, and stage III by 7% with respect to the normal. ($P<0.05$)

Representative brightfield images of a normal and stage III core are presented in Figs. 5 (a) and (b) respectively. Figs. 5 (a') and (b') are the binary versions of the brightfield images obtained through the *Image J* software. They clearly show that the mass density distribution within tissue increases with the progression of cancer. As seen in Fig. 5 (c), the normal Prostate tissue possesses the lowest fractal dimension while the fractal dimensions for each stage of cancer increase with its progression. The actual fractal dimension values for each sample are 1.5737 for the normal, 1.5981 for stage I, 1.6302 for stage II, and 1.6798 for stage III. The percent difference between the normal and stage I samples is 2% normal and stage II samples is 4% and the difference between the normal and stage III is 7%. These results suggest that the fractal dimension of a Prostate TMA sample increases with the progression of cancer. The mass density distribution of a tissue sample increases due to the increased cell replication caused by the presence of cancer, supporting the results obtained (15,21).

## 4. Conclusion

In this paper, we explored the possibility of creating a standardized cancer diagnosis test capable of early, accurate detection based on the fractal dimension analysis of paraffin embedded TMA samples. This was done by microscopic imaging of TMA samples of different stages of Pancreatic, Breast, Colon, and Prostate cancer and calculating their fractal dimensions. The results show that in all the cases of cancer, the fractal dimension of a tissue sample will increase as cancer progresses through the different stages, with similar trend. The increase fractal dimension has universilaty of increase in fractal simension. These results support that the histogram of the grayscale average represents the degree of tissue disorganization which reduces with the progression of carcinogenesis. Therefore, the normal of each cancer tissue has miniscule gray scale while the higher stages of each cancer have a higher grayscale or fractal dimension due to more accumulation of mass density. Here, we were able to distinguish the different stages of Pancreatic, Breast, Colon, and Prostate cancer via quantitative approach of the fractal dimension of commercially available TMA samples using optical transmission imaging. This result shows that there is definite potential for the creation of a standardized, quantitative cancer diagnosis approach which is free of some of the issues plaguing the current cancer diagnosis methods. Lastly, an easy, prompt, cost-effective, and accurate quantitative cancer diagnosis method can also be compiled into a database with the possibility for the creation of a shared database accessible to every physician. By compiling the fractal dimensions that are calculated by doctors all over the country, a master list can be created that gives the range of the average fractal dimension for each type and stage of cancer and helps to distinguish the early stages of cancer. This master list would be increasingly more accurate as more data is entered and further improve this early and effective cancer diagnosis method in future.

**Acknowledgments**

The work is partially supported by the Mississippi State University.